\documentclass[twocolumn, pr]{revtex4}
\usepackage{graphicx}
\usepackage{amsmath,amsbsy,amssymb}
\usepackage{bm}
\usepackage{mathrsfs,color}

\newcommand{\diff}{\mathrm{d}}

\newcommand{\imu}{\mathrm{i}}
\newcommand{\epn}{\mathrm{e}}

\newcommand{\ua}{\uparrow}
\newcommand{\da}{\downarrow}
\newcommand{\dg}{\dagger}
\newcommand{\la}{\langle}
\newcommand{\ra}{\rangle}
\newcommand{\al}{\alpha}
\newcommand{\sg}{\sigma}
\newcommand{\gm}{\gamma}
\newcommand{\ep}{\varepsilon}

\begin{document}

\title{
Electronic orders in multi-orbital Hubbard models with lifted orbital degeneracy
}

\author{Shintaro Hoshino$^{1}$ and Philipp Werner$^2$}

\affiliation{
$^1$Department of Basic Science, The University of Tokyo, Meguro, Tokyo 153-8902, Japan
\\
$^2$Department of Physics, University of Fribourg, 1700 Fribourg, Switzerland
}

\date{\today}

\begin{abstract}
We study the symmetry-broken phases in two- and three-orbital Hubbard models with lifted orbital degeneracy using dynamical mean field theory.
On the technical level, we explain how symmetry relations can be exploited to measure the four-point correlation functions needed for the calculation of the lattice susceptibilities. In the half-filled two-orbital model with crystal field splitting, we find an instability of the metallic phase to spin-orbital order with neither spin nor orbital moment. This ordered phase is shown to be related to the recently discovered fluctuating-moment induced spin-triplet superconducting state in the orbitally degenerate model with shifted chemical potential. In the three-orbital case, we consider the effect of a crystal field splitting on 
the spin-triplet superconducting state in the model with positive Hund coupling, and the spin-singlet superconducting state in the case of negative Hund coupling.
It is demonstrated that
for certain crystal field splittings 
the higher energy orbitals instead of the lower ones are relevant for superconductivity, 
and that $T_c$ can be enhanced by the crystal field effect.  
We comment on the implications of our results for the superconductivity in 
strontium ruthenates, and for the recently reported light-enhanced superconducting state in 
alkali-doped fullerides.
\end{abstract}


\maketitle

\section{Introduction}

The local Slater-Kanamori interaction \cite{slater19xx,kanamori1963} originating from Coulomb repulsion leads to highly nontrivial phase diagrams and crossover phenomena in multi-orbital Hubbard models. Depending on the filling and the energy splittings between the orbitals one finds antiferromagnetic or ferromagnetic order \cite{chan2009,hoshino2015}, orbital order \cite{chan2009}, high-spin/low-spin transitions \cite{werner2007,suzuki2009}, staggered high-spin/low-spin order \cite{kunes2014}, excitonic insulating phases \cite{kunes2014b,kunes2015}, or intra-orbital spin-triplet (equal-spin) pairing \cite{hoshino2015}. Some of these instabilities are linked to the spin-freezing crossover \cite{werner2008}, which occurs in models with nonzero (positive) Hund coupling, and underlies the unusual finite-temperature properties of Hund metals \cite{georges2013}. In models with negative Hund coupling, an intra-orbital spin-singlet superconducting state appears \cite{capone2009,koga2015}, but alternative ordered states have not yet been systematically explored.

An unbiased way of mapping out the 
electronic instabilities to long-range orders 
is to compute the corresponding susceptibilities and to look for divergences as a function of the model parameters. While such calculations cannot be easily performed with numerically exact lattice methods, they become computationally tractable within the dynamical mean field theory (DMFT) approximation \cite{georges1996}. This theory assumes a spatially local self-energy and vertex, and produces qualitatively correct solutions for high-dimensional lattice models. The use of 
a featureless density of states in DMFT ensures that the results do not depend on subtle bandstructure effects. In this work, we employ the DMFT formalism and a semi-circular density of states to map out the  instabilities to 
uniform and staggered long-range ordered phases in two and three orbital systems, focusing on the physically most interesting intermediate coupling regime.

One main purpose of this study is to show how symmetry relations can be exploited in the numerical simulations based on continuous-time Monte Carlo impurity solvers \cite{gull2011}, and in the analysis of the ordering instabilities. In particular, we will demonstrate that the spin-orbital ordered phase 
(also called excitonic insulator phase \cite{kunes2014b,kunes2015}) 
appearing in half-filled two-orbital models with crystal field (CF) splitting is related by symmetry operations to the orbital-singlet spin-triplet superconducting state found in the orbitally degenerate model away from half-filling \cite{klejnberg1999, spalek2001, han2004, sakai2004, kubo2007, zegrodnik2014}.
We also use symmetry relations to analyze and explain the Cooper pair formation in three orbital systems with lifted orbital degeneracy.

For positive (ferromagnetic) Hund coupling, the Slater-Kanamori interaction describes t$_{\rm 2g}$-based three orbital systems, as realized for example in Sr$_2$RuO$_4$ and SrRuO$_3$ \cite{georges2013}. These materials indeed exhibit spin-triplet superconductivity or ferromagnetism, as well as the non-Fermi liquid properties associated with spin-freezing. 
Recently the authors have demonstrated that the spin-triplet pairing 
is induced by the fluctuating local moments which appear at the border of the spin-frozen regime 
and eventually order ferromagnetically or antiferromagnetically \cite{hoshino2015}.
Similar multi-orbital physics is also expected to be relevant for other types of spin-triplet pairing, such as realized in U-based ferromagnetic superconductors \cite{aoki2012}.
In this paper we will examine the effect of CF splittings on the spin-triplet pairing.
We will show that 
the Cooper pairs can be formed in the higher-energy lifted orbitals instead of lower one. 
This somewhat counter-intuitive result can be explained by exploiting a particle-hole transformation.

The half-filled three-orbital model with negative Hund coupling exhibits an intra-orbital pairing state which is relevant for fulleride superconductors \cite{capone2009,nomura2012,nomura2015}. The effect of CF splittings 
on this superconducting state is of interest in connection with the recently reported light-enhanced superconductivity in K$_3$C$_{60}$ \cite{mitrano2015}. We show that in equilibrium, the lifting of the orbital degeneracy can slightly stabilize 
the pairing, but the effect on $T_c$ is small.
Hence, it is unlikely that the experimentally observed dramatic enhancement of $T_c$
can be explained by quasi-static distortions of the C$_{60}$ molecules within an equilibrium picture.

The outline of the paper is as follows. In Sec.~II, we present the model and explain its basic properties, in particular the symmetries. 
We discuss in Sec.~III some technical details concerning the calculation of lattice susceptibilities within DMFT. Section~IV shows the results for the two-orbital model, and explains the connection between spin-orbital order and spin-triplet superconductivity, while Sec.~V presents results for the three-orbital model with lifted orbital degeneracy. Section~VI is a brief summary and conclusion.

\section{Model and Method}
We consider the Hubbard model with $M$ degenerate orbitals given by the general Hamiltonian 
\begin{align}
&\mathscr{H} = \sum_{\bm k\gm\sg} (\ep_{\bm k}-\mu) c^\dg_{\bm k\gm\sg} c_{\bm k\gm \sg}
+ U \sum_{i\gm} n_{i\gm\ua} n_{i\gm\da}
\nonumber \\
&\ 
+ U' \sum_{i\sg,\gm<\gm'} n_{i\gm\sg} n_{i\gm'\bar \sg}
+ (U'-J) \sum_{i\sg,\gm<\gm'} n_{i\gm\sg} n_{i\gm'\sg}
\nonumber \\
&\ 
- \alpha J \sum_{i,\gm<\gm'} (
 c^\dg_{i\gm\ua} c_{i\gm\da} c^\dg_{i\gm'\da} c_{i\gm'\ua}
+c^\dg_{i\gm\ua} c^\dg_{i\gm\da} c_{i\gm'\ua} c_{i\gm'\da}
+{\rm H.c.} ). 
\label{hamilt}
\end{align}
The operator $c_{i\gm\sg}$ annihilates the electron with orbital $\gm$ and spin $\sg$ at site $i$.
The Fourier transformation is defined by $c_{\bm k\gm\sg} = N^{-1/2} \sum_i c_{i\gm\sg} \epn^{\imu\bm k \cdot \bm R_i}$, where $N$ is the number of sites and $\bm R_i$ the spatial coordinate at site $i$.
The number operator is defined by $n_{i\gm\sg} = c_{i\gm\sg}^\dg c_{i\gm\sg}$.
In Eq.~(\ref{hamilt}), the parameter $\alpha$ is introduced to describe the effect of spin anisotropy
and orbital anisotropy. 
Physically this anisotropy originates from the spin-orbit coupling.
We note that for repulsively interacting systems, the pair hopping term is irrelevant.
This is because the strongest intra-orbital repulsion $U$ disfavors the doubly occupied orbital state and as a result the pair hopping process between orbitals rarely occurs. 
If the Hund coupling $J$ is negative, as in models discussed in connection with fulleride superconductors \cite{capone2009,nomura2015}, the pair hopping becomes relevant, while spin-flip processes are effectively suppressed.
With an anisotropy parameter $\al < 1$, the effect of these spin-flip and pair-hopping process becomes weaker.

Let us first consider the 
isotropic case with $\alpha=1$.
In rotationally invariant systems, where the relation $U=U'+2J$ holds, the interaction part of the Hamiltonian can be rewritten in the form 
\begin{align}
&\mathscr{H}_{\rm int} =
 \sum_{i\gm\gm'\sg\sg'}
\left[
 \frac{U'}{2} c^\dg_{i\gm\sg} c_{i\gm\sg} c^\dg_{i\gm'\sg'} c_{i\gm'\sg'}
\right.
\nonumber \\
& \hspace{2mm}
\left.
+\frac{J}{2}  c^\dg_{i\gm\sg} c_{i\gm'\sg} c^\dg_{i\gm'\sg'} c_{i\gm\sg'}
+\frac{J}{2}  c^\dg_{i\gm\sg} c_{i\gm'\sg} c^\dg_{i\gm\sg'}  c_{i\gm'\sg'}
\right],
\label{hamilt2}
\end{align}
which is suitable for discussing the symmetries of this model.
Using this expression, we can easily verify that the transformation
\begin{align}
\mathscr{S} c_{i\gm\sg} \mathscr{S}^{-1}
&=\epn^{\imu\theta} \sum_{\sg'} {\cal U}_{\sg\sg'} \sum_{\gm'} {\cal V}_{\gm\gm'} c_{i\gm'\sg'} 
\end{align} 
does not change $\mathscr{H}_{\rm int}$, 
if $\cal{U}$ and $\cal{V}$ are $2\times 2$ unitary and $M\times M$ orthogonal matrices, respectively. 
This invariance implies an U(1)$\times$SU(2)$\times$SO($M$) symmetry.

On the other hand, in the anisotropic case with $\al=0$, there are only density-density type interactions.
Then the symmetry is described by $[U(1)]^{2M}$ where $2M$ is the number of spin/orbital indices.
The relevant symmetry operation is given 
by ${\mathscr{S}}c_{i\gm\sg} {\mathscr{S}}^{-1} = \epn^{\imu\theta_{\gm\sg} } c_{i\gm\sg}$ for any $(\gm,\sg)$.

When we consider the degenerate-orbital model including kinetic terms, the total Hamiltonian is also unchanged by this local transformation.
In this case we have the relation
\begin{align}
\la \mathscr{O} \ra = \la \mathscr{S} \mathscr{O} \mathscr{S}^{-1} \ra
\label{symmetry_relation}
\end{align}
for arbitrary operators $\mathscr{O}$.
This relation will be used in Sec.~IIIC to 
derive measurement formulas for 
the two-particle Green functions.

We analyze the multi-orbital Hubbard model within the framework of DMFT \cite{georges1996}. 
This theory becomes exact in the limit of infinite dimensions, where only local correlations are relevant.
Because of the local self-energy, the lattice problem can be mapped onto a multi-orbital impurity problem with local interaction terms identical to those of Eq.~(\ref{hamilt}) and noninteracting baths whose properties can be encoded by the hybridization functions $\Delta_{\gamma\sigma}$.
We consider the infinite-dimensional Bethe lattice, 
whose non-interacting density of states has a semi-circular shape: $\rho(\ep) = (1/2\pi t^2)\sqrt{4t^2-\ep^2}$ with $t=1$ the (rescaled) hopping integral.
For this lattice, the DMFT self-consistency condition simplifies to $\Delta_{\gamma\sigma}=t^2 G_{\gamma\sigma}$, with $G_{\gamma\sigma}$ the impurity Green function.
Since the wave vector in the Bethe lattice is ill-defined,
it is natural to work in a  real space representation.
On the other hand we can also consider an alternative description in terms of ``pseudo-wave-vectors" (see Appendix~\ref{appendixBethe}), 
which enables a discussion analogous to that for ordinary periodic lattices.

\section{Evaluation of Susceptibilities}

\subsection{Definition}

In this paper we discuss the instabilities toward long-ranged orders.
We consider
ordered states corresponding to operators 
\begin{align}
\mathscr{O}(\bm q) = \sum_i \mathscr{O}_i \epn^{-\imu \bm q\cdot \bm R_i},
\end{align}
which can be detected by the divergence of the susceptibilities defined by
\begin{align}
\chi_{\mathscr{O}} (\bm q) = \frac 1 N \int _{0} ^\beta \la \mathscr{O}(\bm q, \tau) \mathscr{O}(-\bm q) \ra  \, \diff \tau.
\label{eq:suscep_def}
\end{align}
Here $\mathscr{O}(\tau) = \epn^{\tau\mathscr{H}} \mathscr{O} \epn^{-\tau \mathscr{H}}$ is the Heisenberg picture with imaginary time $\tau$.
In DMFT, these susceptibilities can be calculated from the local vertex parts extracted from the effective impurity problem \cite{georges1996}.

Let us consider the specific forms of $\mathscr{O}$ by taking the two-orbital model as an example.
For diagonal orders, the 
operators 
$\mathscr{O}_i$ 
are given by
\begin{align}
n_i &= \sum_{\gm\sg}c^\dg_{i\gm\sg} c_{i\gm\sg},
\label{eq:op_charge} \\
s^\mu_i &= \sum_{\gm\sg\sg'}c^\dg_{i\gm\sg} \sg^{\mu}_{\sg\sg'} c_{i\gm\sg'},
\label{eq:op_spin} \\
\tau^\nu_i &= \sum_{\gm\gm'\sg}c^\dg_{i\gm\sg} \sg^{\nu}_{\gm\gm'} c_{i\gm'\sg},
\label{eq:op_orbital} \\
o^{\nu\mu}_i &= \sum_{\gm\gm'\sg\sg'}c^\dg_{i\gm\sg} \sg^{\nu}_{\gm\gm'} \sg^{\mu}_{\sg\sg'} c_{i\gm'\sg'},
\label{eq:op_spinorbital}
\end{align}
corresponding to charge, spin, orbital and spin-orbital moments, respectively.
Here $\mu,\nu=x,y,z$.
For offdiagonal orders, we have
\begin{align}
p^{{\rm s}\mu}_i &= \tfrac{1}{2} \sum_{\gm\gm'\sg\sg'}c^\dg_{i\gm\sg} \epsilon_{\gm\gm'} (\sg^\mu \epsilon)_{\sg\sg'} c^\dg_{i\gm'\sg'} + {\rm H.c.},
\label{eq:pair1} \\
p^{\nu{\rm s}}_i &= \tfrac{1}{2} \sum_{\gm\gm'\sg\sg'}c^\dg_{i\gm\sg} (\sg^\nu \epsilon)_{\gm\gm'} \epsilon_{\sg\sg'}  c^\dg_{i\gm'\sg'} + {\rm H.c.}, 
\label{eq:pair2}
\end{align}
which correspond to the orbital-singlet-spin-triplet and orbital-triplet-spin-singlet pairing amplitudes, respectively. The anti-symmetric unit tensor is defined by $\epsilon = \imu \sg^y$.
All of these operators are hermitian.

To further classify the diagonal operators given above, we consider the time-reversal operation defined by $\mathscr{T} = \exp(-\imu\pi\sum_i s^y_i/2) \mathscr{K}$ with complex conjugation operator $\mathscr{K}$.
This transforms the electron operator as
\begin{align}
\mathscr{T} c_{i\gm\sg} \mathscr{T}^{-1} = \sum_{\sg'} \epsilon_{\sg\sg'} c_{i\gm\sg'}.
\end{align}
Note that $\mathscr{T}$ is an 
antiunitary operator: 
$\mathscr{T} z \mathscr{T}^{-1}=z^*$ for a complex number $z$.
This operation 
changes the momentum ($\bm k\rightarrow -\bm k$) and flips the spin state ($\ua\rightarrow \da$ or $\da\rightarrow \ua$).
As expected, the charge is time-reversal even ($\mathscr{T} n_i \mathscr{T}^{-1}=n_i$) and spin is odd ($\mathscr{T} s_i^\mu \mathscr{T}^{-1}=-s_i^\mu$).
The orbital moments $\tau^x_i$ and $\tau^z_i$ are time-reversal even, while $\tau^y_i$ is odd.
The difference between the orbital moments arises due to the presence of the imaginary unit in the Pauli matrix $\sg^y$, which gives an additional minus sign under the time-reversal operation.

The results are summarized in Tab. \ref{tab:time-reversal}.
While in this paper we focus on a model study, a physical interpretation in terms of the doubly degenerate e$_{\rm g}$ orbitals is given in Appendix~\ref{appendix_eg}, where the operator $\tau^y$ is shown to be a (generalized) magnetic moment.
These arguments can also be applied to the three orbital model, where the Pauli matrix for orbital is replaced by the Gell-Mann matrix.
In this case we have eight kinds of orbital orders, three of which are time-reversal odd operators.

For the pairing state characterized by Eqs.~\eqref{eq:pair1} and \eqref{eq:pair2}, the time-reversal symmetry is not broken.
More specifically, while the time-reversal operation $\mathscr{T}$ can change the sign of these operators, this can be absorbed by simultaneously performing a global U(1) gauge transformation. 
On the other hand, if two or more Cooper pairs become finite with different phases, then the resulting state has broken time-reversal symmetry \cite{sigrist1991}.
In this case the sign cannot be absorbed by the gauge transformation, and thus the time-reversal symmetry is broken.

\begin{table}[t]
\begin{tabular}{c|c|c}
\hline
identifier & operator  & time-reversal 
\\ 
\hline
(i)&$n$ & even 
\\
(ii)&$s^x$,$s^y$,$s^z$ & odd 
\\
(iii)&$\tau^x,\tau^z$ & even 
\\
(iv)&$\tau^y$ & odd 
\\
(v)&$o^{xx},o^{xy},o^{xz},o^{zx},o^{zy},o^{zz}$ & odd 
\\
(vi)&$o^{yx},o^{yy},o^{yz}$ & even 
\\
(vii)&$p^{sx},p^{sy},p^{sz}$ & even 
\\
(viii)&$p^{xs},p^{zs}$ & even 
\\
(ix)&$p^{ys}$ & even 
\\
\hline
\end{tabular}
\caption{
Classification of operators relevant to diagonal orders for the rotationally invariant case ($\al=1$).
}
\label{tab:time-reversal}
\end{table}

\subsection{Two-particle Green function and susceptibilities}

Here we discuss the calculation of the susceptibility in the framework of DMFT \cite{georges1996}.
For diagonal and offdiagonal orders, the relevant two-particle Green functions are given by
\begin{align}
&\chi^{\rm diag}_{ij,a_1a_2a_3a_4} (\tau_1, \tau_2, \tau_3, \tau_4) =
\nonumber \\
&\hspace{8mm}\la T_\tau c^\dg_{ia_1} (\tau_1) c_{ia_2} (\tau_2) c^\dg_{ja_3} (\tau_3) c_{ja_4} (\tau_4) \ra
\nonumber \\
&\hspace{8mm}-
\la T_\tau c^\dg_{ia_1} (\tau_1) c_{ia_2} (\tau_2) \ra \la T_\tau c^\dg_{ja_3} (\tau_3) c_{ja_4} (\tau_4) \ra
, \label{eqn:tpgf}
\\
&\chi^{\rm offd}_{ij,a_1a_2a_3a_4} (\tau_1, \tau_2, \tau_3, \tau_4) =
\nonumber \\
&\hspace{8mm}\la T_\tau c^\dg_{ia_1} (\tau_1) c^\dg_{ia_2} (\tau_2) c_{ja_3} (\tau_3) c_{ja_4} (\tau_4) \ra
, \label{eqn:tpgf2}
\end{align}
respectively.
The index $a$ denotes the spin and orbital indices $(\gm,\sg)$.
Here $T_\tau$ represents the imaginary time-ordering operator, and we subtract the disconnected part in Eq.~\eqref{eqn:tpgf}.
We also define the Fourier transformation with respect to imaginary time by
\begin{align}
&\chi^\xi_{ij,a_1a_2a_3a_4} (\imu \ep_n, \imu \ep_{n'})
=
\nonumber \\
&\hspace{8mm}\frac{1}{\beta^2} \int_0^\beta \diff\tau_1\diff\tau_2\diff\tau_3\diff\tau_4
\, \chi^\xi_{ij,a_1a_2a_3a_4} 
(\tau_1, \tau_2, \tau_3, \tau_4) 
\nonumber \\
&\hspace{39mm}\times \epn^{\imu \ep_n (\tau_2 - \tau_1) }
\, \epn^{\imu \ep_{n'} (\tau_4 - \tau_3)}
,
\end{align}
where $\xi$ means `diag' or `offd'.
Since we are interested only in static susceptibilities, we set the bosonic frequency to zero.
The susceptibilities defined in Eq.~\eqref{eq:suscep_def} can be derived by summing up the fermionic Matsubara frequencies in the two-particle Green function, which is explicitly written in the form
\begin{align}
\chi^\xi_{\eta}(\bm q) &= \frac{T}{N}\sum_{nn'}\sum_{ij}\sum_{a_1a_2a_3a_4}{\cal A}^\eta_{a_1a_2}{\cal A}^\eta_{a_4a_3} \epn^{-\imu \bm q\cdot (\bm R_i - \bm R_j)}
\nonumber \\
&\ \ \ \ \ 
\times\chi^\xi_{ij, a_1a_2a_3a_4} (\imu\ep_n, \imu\ep_{n'})
,
\end{align}
where the ${\cal A}^\eta_{aa'}$ are form factors originating from Eqs.~(\ref{eq:op_charge}--\ref{eq:op_spinorbital}) for $\xi=$`diag', and from Eqs.~(\ref{eq:pair1},\ref{eq:pair2}) for $\xi=$`offd'.

Now we employ the Bethe-Salpeter equation which relates the two-particle Green function and vertex. 
Since the vertex part $\Gamma$ is local in DMFT \cite{georges1996}, it can be evaluated from the local two-particle Green function as
\begin{align}
&\chi^\xi_{ii,a_1a_2a_3a_4} (\imu \ep_n, \imu \ep_{n'})
=
\chi_{ii,a_1a_2a_3a_4}^{\xi 0} (\imu \ep_n, \imu \ep_{n'})
\nonumber \\
&+
\sum_{n_1n_2}
\sum_{aa'a''a'''}
\chi_{ii,a_1a_2aa'}^{\xi0} (\imu \ep_n, \imu \ep_{n_1})
\Gamma_{i,a'aa'''a''} (\imu \ep_{n_1}, \imu \ep_{n_2})
\nonumber \\
&\hspace{20mm}\times
\chi^\xi_{ii,a''a'''a_3a_4} (\imu \ep_{n_2}, \imu \ep_{n'})
. \label{eqn:bs_eq}
\end{align}
Here the two-particle Green function without vertex parts is written as $\chi^{\xi 0}$.
The site index for the vertex part can be neglected if the system is uniform.

The local vertex extracted above is inserted into the non-local Bethe-Salpeter equation 
\begin{align}
&\chi^{\xi}_{ij,a_1a_2a_3a_4} (\imu \ep_n, \imu \ep_{n'})
=
\chi_{ij,a_1a_2a_3a_4}^{\xi 0} (\imu \ep_n, \imu \ep_{n'})
\nonumber \\
&+
\sum_{\ell n_1n_2}
\sum_{aa'a''a'''}
\chi_{i\ell,a_1a_2aa'}^{\xi 0} (\imu \ep_n, \imu \ep_{n_1})
\Gamma_{\ell,a'aa'''a''} (\imu \ep_{n_1}, \imu \ep_{n_2})
\nonumber \\
&\hspace{22mm}\times
\chi^{\xi}_{\ell j,a''a'''a_3a_4} (\imu \ep_{n_2}, \imu \ep_{n'}).
\end{align}
In practice, it is convenient to perform the Fourier transformation with respect to the site index before solving the matrix equation.
Thus we obtain the two-particle lattice Green functions, which contain the information of the susceptibilities.

\subsection{Measurement trick for local two-particle Green functions}

As explained in the previous section, the local vertices can be extracted from the local two-particle Green functions.
In DMFT calculations of the multi-orbital Hubbard model \eqref{hamilt}, 
we must consider the following two-particle Green functions of the corresponding impurity model:
\begin{align}
\chi^{\rm a}_{\gm\gm'\sg\sg'} &= \la T_\tau c^\dg_{\gm\sg}(\tau_1) c_{\gm\sg}(\tau_2) c^\dg_{\gm'\sg'}(\tau_3) c_{\gm'\sg'}(\tau_4) \ra,
\label{eq:chi_a} \\
\chi^{\rm b}_{\gm\gm'\sg\sg'} &= \la T_\tau c^\dg_{\gm\sg}(\tau_1) c_{\gm\sg'}(\tau_2) c^\dg_{\gm'\sg'}(\tau_3) c_{\gm'\sg}(\tau_4) \ra,
\label{eq:chi_b} \\
\chi^{\rm c}_{\gm\gm'\sg\sg'} &= \la T_\tau c^\dg_{\gm\sg}(\tau_1) c_{\gm'\sg}(\tau_2) c^\dg_{\gm\sg'}(\tau_3) c_{\gm'\sg'}(\tau_4) \ra.
\label{eq:chi_c} 
\end{align}
All of the diagonal and offdiagonal two-particle Green functions that appear in the previous subsection can be calculated from $\chi^{\rm a,b,c}$. 
A powerful method to solve impurity models is the continuous-time Monte Carlo technique \cite{gull2011}. 
Here, we employ the hybridization expansion method in the matrix formulation \cite{werner2006_matrix}.  
In this approach, the function $\chi^{\rm a}$ can be calculated by the standard procedure which involves the removal of two hybridization functions \cite{gull2011}.  
However, the other functions 
$\chi^{\rm b,c}$, 
which are in general finite,  
cannot be obtained by the standard technique unless the impurity model has  
offdiagonal 
hybridizations.

If we have the continuous SU(2)$\times$SO($M$) symmetry the two-particle Green functions $\chi^{\rm b}$ and $\chi^{\rm c}$ can be expressed in terms of $\chi^{\rm a}$ using Eq.~\eqref{symmetry_relation}.
To see this, we consider the function $\chi^{\rm b}_{12\ua\da}$ in the two-orbital model ($M=2$).
The SU(2) and SO(2) matrices in the spin and orbital spaces are given by
\begin{align}
\cal{U} = 
\begin{pmatrix}
z & -w^* \\
w & z^*
\end{pmatrix}
,\ \ 
\cal{V} = 
\begin{pmatrix}
x & -y \\
y & x
\end{pmatrix}
, 
\end{align}
where $|z|^2 + |w|^2=1$ ($z,w \in \mathbb{C}$) and $x^2+y^2=1$ ($x,y \in \mathbb{R}$).
The SO(2) matrix in orbital space represents a rotation around the $\tau^y$ axis.
The quantity $\chi^{\rm b}_{12\ua\da}$ is transformed by $\cal{U}$ as
\begin{align}
\chi^{\rm b}_{12\ua\da} \longrightarrow 
(|z|^4 + |w|^4)\chi^{\rm b}_{12\ua\da} + 2|z|^2|w|^2(\chi^{\rm a}_{12\ua\ua} - \chi^{\rm a}_{12\ua\da})
, \label{eq:two_orb_demo}
\end{align}
where we have used the equivalence between $\gm=1,2$ and between $\sg=\ua,\da$.
If the transformation does not change the whole Hamiltonian, which implies $\al=1$, the left- and right-hand sides of Eq.~\eqref{eq:two_orb_demo} can be regarded as identical.
Thus we obtain the relation
\begin{align}
\chi^{\rm b}_{12\ua\da} = \chi^{\rm a}_{12\ua\ua} - \chi^{\rm a}_{12\ua\da}.
\end{align}
Note that this result is not dependent on the values of $z$ and $w$.
This relation has been previously used in the study of a two-channel Kondo lattice \cite{hoshino2014}.  
In a similar manner, if the system has orbital SO(2) symmetry, it satisfies the additional relation
\begin{align}
\chi^{\rm c}_{12\ua\da} = \chi^{\rm a}_{11\ua\da} - \chi^{\rm a}_{12\ua\da} + \tilde \chi^{\rm b}_{12\ua\da},
\end{align}
where we have defined 
the time-sorted quantity
\begin{align}
\tilde \chi
(\tau_1, \tau_2,\tau_3,\tau_4) = 
- \chi
(\tau_1, \tau_4,\tau_3,\tau_2). 
\end{align}
By an analogous trick we can also obtain the relation
\begin{align}
\chi^{\rm c}_{12\ua\ua} = \chi^{\rm a}_{11\ua\ua} - \chi^{\rm a}_{12\ua\ua} + \tilde \chi^{\rm a}_{12\ua\ua}
.
\end{align}
The other types of $\chi^{\rm b}$ and $\chi^{\rm c}$ can be represented by the quantities obtained above.
Thus, for the calculation of the two-particle Green functions $\chi^{\rm a,b,c}$, it is sufficient to measure $\chi^{\rm a}$.

A key prerequisite for this technique is the 
SU(2) or SO(2) symmetry in the Hamiltonian.
Hence, these formulae do not hold when the symmetry is lowered by, e.g., a CF splitting. 
The generation of new terms on the right-hand side of Eq.~\eqref{eq:two_orb_demo} 
is also an important ingredient for the present trick.
For example the U(1) transformation only changes the phase factor,
which is not enough to derive expressions for $\chi^{\rm b}$ and $\chi^{\rm c}$.
Thus, at least a two-dimensional representation of the symmetry operation is necessary.

\begin{figure}
\begin{center}
\includegraphics[width=80mm]{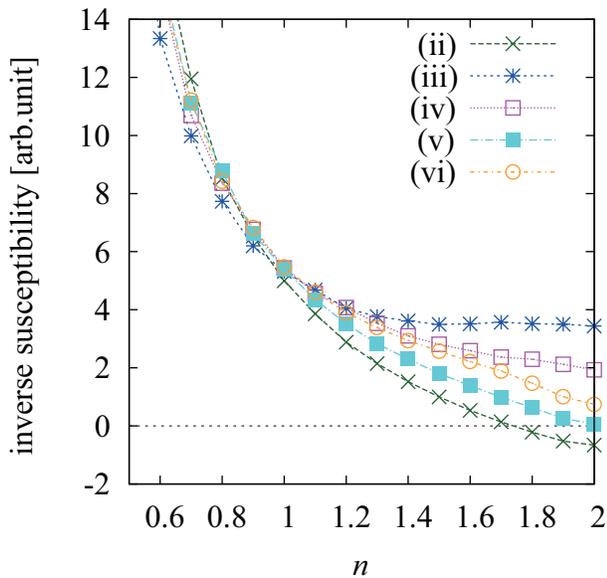}
\caption{(Color online)
Filling dependence of inverse staggered susceptibilities for diagonal orderings in the isotropic case ($\al=1$).
The orders (ii)-(vi) are defined in Tab.~\ref{tab:time-reversal}.   
We choose the parameters $U=2$, $J/U=1/4$ and $T=0.02$.
The staggered charge susceptibility [(i) in Tab.~\ref{tab:time-reversal}] is not shown here because of low numerical accuracy.
}
\label{fig:suscep}
\end{center}
\end{figure}

Let us also add a comment on the 
phase transformation.
While usually it changes the phase of the fermion operators uniformly
($c_{i\gm\sg} \rightarrow c_{i\gm\sg}\epn^{\imu \theta}$), here we can slightly generalize it as 
$c_{i\gm\sg} \rightarrow c_{i\gm\sg}\epn^{\imu (\theta_\sg + \theta_\gm)}$ because of the SU(2)$\times$SO(2) symmetry.
For the invariance of the Hamiltonian, the phases $\theta_\ua$ and $\theta_\da$ are arbitrary, but $\theta_1 - \theta_2 = \pi m$ ($m\in \mathbb{Z}$) must be satisfied due to the pair-hopping term.
These are related to conservation laws of 
spin/orbital indices.
Combining this transformation with Eq.~\eqref{symmetry_relation}, it can be explicitly shown that expectation values such as 
$\la c^\dg_{1\ua}c_{1\da}c^\dg_{1\ua}c_{1\da} \ra$ and $\la c^\dg_{1\ua}c_{1\ua}c^\dg_{1\ua}c_{2\ua} \ra$ are identically zero.

The above techniques can also be used in models with a general number $M$ of orbitals.
Namely, if we choose two of the $M$ orbitals, we can utilize the partial SO(2) symmetry within this subspace, so that the procedure remains unchanged.

\section{Two-orbital model}

\subsection{Degenerate orbitals}

In this section, we study the ordering instabilities in the two-orbital model with $J>0$ and first consider the orbitally degenerate case (Eq.~(\ref{hamilt})). Figure~\ref{fig:suscep} shows the staggered susceptibilities characterized by the ``ordering vector $\bm Q$'' (see Appendix~\ref{appendixBethe}) in the spin-isotropic system with $\al=1$.
Here the symmetry tricks explained in the previous section are used for the evaluation of the two-particle Green functions.
The uniform susceptibilities for diagonal and offdiagonal orders do not diverge for the chosen parameters, and therefore are not shown.

Near half filling ($n\lesssim 2$), the antiferromagnetic order with $\bm s(\bm Q)$ appears as demonstrated by the sign change of the corresponding inverse susceptibility in Fig.~\ref{fig:suscep}.
The other susceptibilities remain positive and hence no further instabilities are present.
For the low-filling case with $n\lesssim 1$, all of the susceptibilities behave in a similar manner.
This is because the interaction effect is weaker and an approximate SU(4) symmetry appears.
For the present choice of parameters no other symmetry breaking, including superconductivity, occurs.

\begin{table}[b]
\begin{tabular}{c|c}
\hline
identifier & operator
\\ 
\hline
(i')&$n$
\\
(ii')&$s^z$
\\
(iii')&$s^x$,$s^y,o^{zx},o^{zy}$
\\
(iv')&$\tau^z$
\\
(v')&$\tau^x,\tau^y,o^{xz},o^{yz}$ 
\\
(vi')&$o^{zz}$ 
\\
(vii')&$o^{xx},o^{xy},o^{yx},o^{yy}$ 
\\
(viii')&$p^{sx},p^{sy}$
\\
(ix')&$p^{zs},p^{sz}$
\\
(x')&$p^{xs},p^{ys}$
\\
\hline
\end{tabular}
\caption{
Classification of independent operators for the Ising-anisotropic case ($\al=0$) without CF splitting.
}
\label{tab:ising}
\end{table}

In our study, although the survey is limited due to the high numerical cost for $\al\neq 0$, we did not find any interesting ordered states for the spin isotropic case ($\al=1$) with $J>0$. 
In the following, we therefore focus on the model with anisotropic interaction ($\al=0$).  
The corresponding phase diagram is richer, since superconductivity appears in addition to trivial magnetic orders.
Because the spin-flip and pair-hopping terms in Eq.~\eqref{hamilt2} are dropped in the spin-anisotropic case with $\al=0$, 
we do not have to consider$\chi^{\rm b,c}$ and 
the calculation of the two-particle Green functions simplifies substantially. The independent order parameters are also changed from the previous section, and are listed in Tab.~\ref{tab:ising}. We will consider only uniform ($\bm q=\bm 0$) and staggered ($\bm q=\bm Q$) ordering vectors.

\begin{figure}[t]
\begin{center}
\includegraphics[width=85mm]{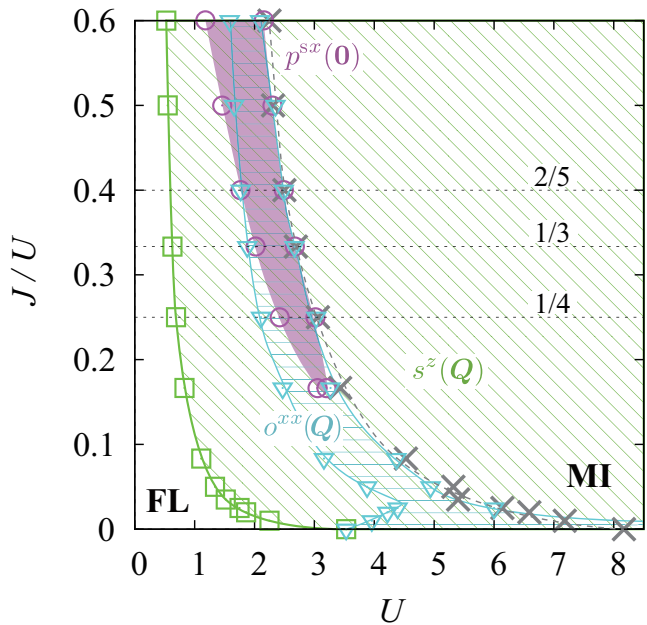}
\caption{(Color online)
Phase boundaries determined by the divergent points of susceptibilities at $n=2$ and $T=0.02$.
The cross symbols indicate Mott transition points. 
The line $J/U=1/4$ is the ratio considered in this paper, $J/U=1/3$ marks the border from repulsive to attractive interactions, and $J/U=2/5$ is a high symmetry line (see text).
The bold letters show Fermi liquid (FL) and Mott insulator (MI) phases without long-range orders.
}
\label{fig:phase_J}
\end{center}
\end{figure}

In the following of this section, we focus on $s^z$, $o^{xx}$, and $p^{{\rm s}x}$, since the corresponding symmetry breaking appears in regions which are not dominated by other ordering, for the parameters considered here. 
Figure~\ref{fig:phase_J} shows the phase diagram in the plane of $U$ and $J/U$ at half filling ($n=2$) and $T=0.02$.
For $J=0$, because of the high symmetry SU(4), all the diagonal orders except for the charge order are degenerate.
On the other hand, a finite $J$ substantially stabilizes the magnetic order with $s^z(\bm Q)$.
At the same time, the critical value of the metal-insulator transition is reduced by the Hund coupling.
There appears staggered spin-orbital order ($o^{xx}(\bm Q)$) and orbital-singlet/spin-triplet pairing ($p^{{\rm s}x}(\bm 0)$), although these ordered regions are covered by the $s^z(\bm Q)$ order. 
The spin-orbital order realized at $J=0$ is destabilized by a small Hund coupling $J$.
This is because the $c^\dg_{1\ua}c_{2\da}$ operator is relevant for the spin-orbital order with $o^{xx}$, and this 
spin-orbital flipping process is suppressed by a positive $J$ which favors equal-spin states.
However, it is stabilized near the Mott transition due to local spin and charge fluctuations. 
As will be shown later, these orders become most stable 
away from half-filling or in the presence of a CF splitting.

It is notable that at $J/U=0.4$ the orders with $o^{xx}(\bm Q)$ and $p^{{\rm s}x}(\bm 0)$ have the same transition point.
This can be understood by symmetry considerations.
We first define the particle-hole transformation for orbital 2, 
\begin{align}
\mathscr{P} c_{i2\sg} \mathscr{P}^{-1} = \sum_{\sg'} \sigma^x_{\sg\sg'} c^\dg_{i2\sg'} \epn^{\imu\bm Q\cdot \bm R_i},
\end{align}
which leaves $c_{i1\sg}$ unchanged. 
This is similar to the transformation from repulsive to attractive interactions in the half-filled single-band Hubbard model \cite{shiba1972}.   
One can show that this transformation does not change the Hamiltonian 
with Ising anisotropy ($\al=0$) at $J/U = 2/5$ and $n=2$.
(In fact, this invariance holds also for the case where the spin-flip term is included, but not in the presence of the pair-hopping term.) 
On the other hand, $\mathscr{P}$ transforms the offdiagonal order into a diagonal order:
\begin{align}
\mathscr{P} p^{{\rm s}x} (\bm 0) \mathscr{P}^{-1}
&= o^{xx} (\bm Q).
\label{Psym}
\end{align}
Thus, these two orders are degenerate at this particular value of $J/U$.

\begin{figure}[t]
\begin{center}
\includegraphics[width=85mm]{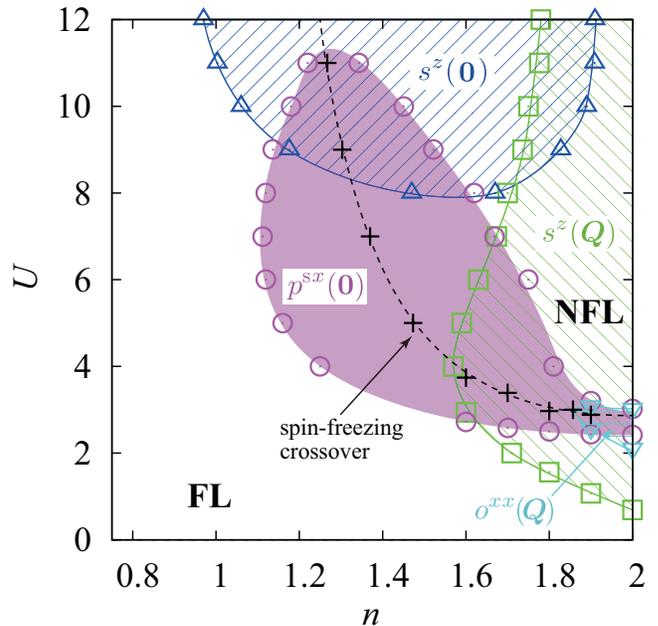}
\caption{(Color online)
Filling ($n$) versus interaction ($U$) phase diagram of the two-orbital model with $J/U=1/4$ at $T=0.02$. 
The Fermi liquid (FL) and non-Fermi liquid (NFL) regimes are separated by the spin-freezing crossover line.
}
\label{fig:away_half}
\end{center}
\end{figure}

Figure~\ref{fig:away_half} plots the phase diagram of the orbitally degenerate model away from half-filling, for $T=0.02$ and $J/U=1/4$. Near half-filling, an antiferromagnetically ordered phase is found, while in the large $U$ regime, a ferromagnetic phase extends over a wide filling range. Between these two magnetically ordered regions, we find an instability toward an orbital-singlet spin-triplet superconducting state with $p^{{\rm s}x}(\bm 0)$. 
Here, the quasi-local Cooper pair results from purely repulsive interactions, by a mechanism that has been proposed in the cold atom context \cite{inaba2012}. 
This phase extends from the end-point of the half-filled Mott insulator along the so-called spin freezing line \cite{werner2008} into the metallic regime, in analogy to the three orbital results discussed in Ref.~\onlinecite{hoshino2015}.
The spin-freezing phenomenon is well characterized by the local quantity 
\begin{align}
\Delta \chi_{\rm loc} = \int_0^\beta d\tau \left[
\la s_i^z (\tau) s_i^z\ra - \la s_i^z(\beta/2) s_i^z \ra
\right]
. \label{eq:def_local_fluct}
\end{align}
The first term is the local spin susceptibility.
The long-time spin correlator, which appears as the second term, reflects the presence of frozen moments.
The difference $\Delta \chi_{\rm loc}$ thus quantifies the local spin fluctuations, whose maxima in parameter space define the spin-freezing crossover line \cite{hoshino2015}.
This line is closely related to the spin-triplet superconductivity as shown in Fig.~\ref{fig:away_half}.
While the main topic of this paper is the effect of CF splitting, Figs.~\ref{fig:phase_J} and \ref{fig:away_half} are important for a deeper understanding of the resulting ordered states under the CF, as discussed in the following.

\subsection{Split orbitals}

We next consider the half-filled two-orbital model with an additional crystal field (CF) splitting given by
\begin{align}
\mathscr{H}_{\rm CF} = \Delta \sum_{i\sg} (n_{i1\sg} - n_{i2\sg}). 
\end{align}
For $\alpha=1$, the metal-insulator phase diagram of this model without ordering has been studied in Ref.~\onlinecite{werner2007}. A qualitatively similar diagram is obtained also for the $\al=0$ case \cite{hafermann2012}. At large $U$ and small $\Delta$, there is a high-spin (Mott) insulating phase with small orbital polarization. For sufficiently large $\Delta$, the system is in a low-spin insulating state with large orbital polarization, which for $U\rightarrow 0$ is adiabatically connected to the band insulator. At large enough $U$ and finite $T$, there is a transition from the high-spin to the low-spin insulator at 
$\Delta\simeq 3J/2$ \cite{georges2013}. At weaker $U$, there exists an intermediate metallic phase, which for large $J/U$ and low $T$ extends along the $\Delta\simeq 3J/2$ line to rather large values of $\Delta$.

\begin{figure}[t]
\begin{center}
\includegraphics[width=85mm]{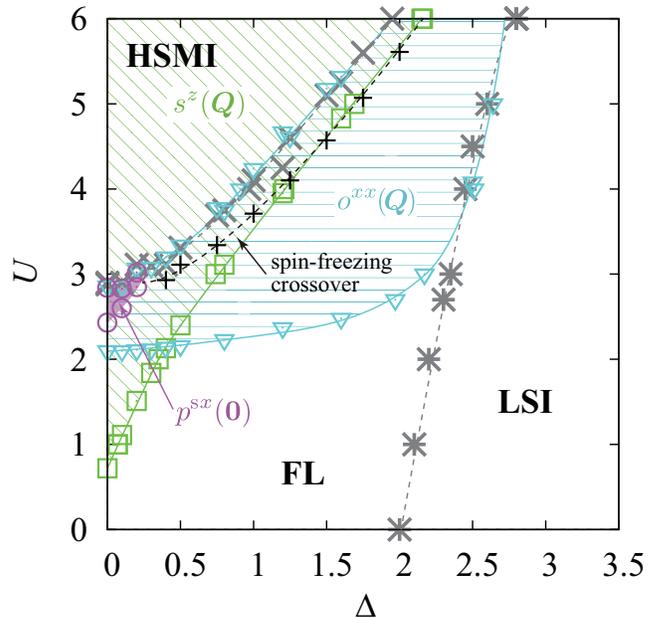}
\caption{(Color online)
Phase diagram of the half-filled two-orbital model with $J/U=1/4$ in the space of CF splitting $\Delta$ and $U$ at temperature $T=0.02$.
In the disordered case, we have Fermi liquid (FL), high-spin Mott insulator (HSMI), and low-spin insulator (LSI) phases.
}
\label{fig:phase_D}
\end{center}
\end{figure}

Figure \ref{fig:phase_D} shows the $\Delta$-$U$ phase diagram for the anisotropic case ($\al=0$) at half filling.
While the high-spin insulator is antiferromagnetically ordered, the metallic nose separating the high-spin and low-spin insulators is unstable to the staggered spin-orbital ordering with $o^{xx}(\bm Q)$. 
This result is consistent with the excitonic insulator state found in this parameter regime in Ref.~\onlinecite{kunes2014b}. The low-spin insulating region is not susceptible to long-range order.
We also find an instability toward the staggered high-spin/low-spin order 
reported in Ref.~\onlinecite{kunes2014}. This instability, which can be detected by the orbital susceptibility with $\tau^z(\bm Q)$, appears for $U\gtrsim 6$. However, we do not show this phase, since it is covered by the $o^{xx}(\bm Q)$ order at least for $U\leq 8$.

We now discuss the connections between the two phase diagrams in Figs.~\ref{fig:away_half} and \ref{fig:phase_D}, which follow from the symmetry relation (\ref{Psym}).  
Away from half-filling, in the model with CF splitting, the spin-triplet superconductivity $p^{{\rm s}x}(\bm 0)$ 
is stabilized along the spin-freezing crossover line, which indicates that local spin-fluctuations induce the pairing state \cite{hoshino2015}.  
If we perform the particle-hole transformation, this pairing state changes into the ordered state with $o^{xx}(\bm Q)$ as discussed above.
At the same time, the chemical potential term 
$\mathscr{H}_{\rm chem}= -\mu \sum_{i\gm\sg} n_{i\gm \sg}$
is transformed into a CF splitting term: 
\begin{equation}
\mathscr{P} \mathscr{H}_{\rm chem} \mathscr{P}^{-1} = - \mathscr{H}_{\rm CF}
\end{equation}
with $\Delta=\mu$. 
In other words, {\it the spin-triplet superconductivity away from half filling corresponds to the spin-orbital order under the CF splitting}. Furthermore the local spin moment is unchanged: $\mathscr{P} s^z_i \mathscr{P}^{-1} = s^z_i$. 
Hence, in both cases the local spin-fluctuations 
play an important role in stabilizing the ordered state. 
As shown in Fig.~\ref{fig:phase_D}, 
there is indeed a spin-freezing line extending along the metallic nose in the region where spin-orbital order appears. 
While this mapping is exact only for $J/U=2/5$, the qualitative correspondence between the two physical situations should be valid in a broader regime, including the case $J/U=1/4$ considered in Figs.~\ref{fig:away_half} and \ref{fig:phase_D}.

\section{Three-orbital model}

\subsection{General remarks}

We are interested in the stability of the inter-orbital-spin-triplet ($J>0$) and intra-orbital-spin-singlet ($J<0$) superconducting phases to CF splittings of the ``2/1'' type 
\begin{align}
\mathscr{H}_{\rm CF} = - \Delta \sum_{i\sg} (n_{i1\sg} + n_{i2\sg} - n_{i3\sg}).
\label{eq:CF_def}
\end{align}
This term breaks the orbital SO(3) symmetry, but the SO(2) symmetry within the $\gm=1,2$ subspace remains for $\al=1$.

\begin{figure}[t]
\begin{center}
\includegraphics[width=85mm]{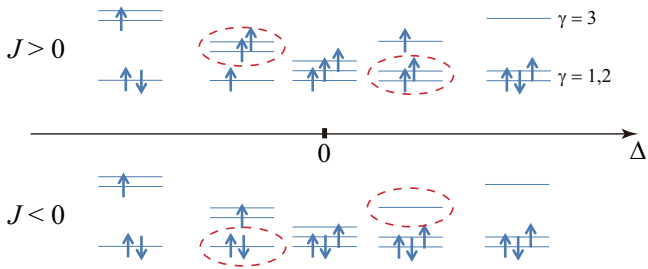}
\caption{(Color online)
Schematic illustration of atomic configurations as a function of $\Delta$ at half-filling ($n=3$). 
The circles with dotted lines show the orbital which is responsible for superconductivity.
For the case with $n=2$, one of three electrons should be removed from figure.
}
\label{fig:config}
\end{center}
\end{figure}

At half-filling and without CF splitting, the Hamiltonian is invariant under the simple particle-hole transformation defined by
\begin{align}
\mathscr{P}_0 c_{i\gm\sg} \mathscr{P}_0^{-1} = c^\dg_{i\gm\sg} \epn^{\imu \bm Q\cdot \bm R_i},
\end{align}
i.e., we have $\mathscr{P}_0 \mathscr{H} \mathscr{P}_0^{-1} = \mathscr{H}$.
On the other hand, the CF Hamiltonian is transformed as $\mathscr{P}_0 \mathscr{H}_{\rm CF} \mathscr{P}_0^{-1} = - \mathscr{H}_{\rm CF}$.
Thus the sign of the CF splitting $\Delta$ is reversed.
For some interaction and filling parameters, as seen below, the present system shows superconductivity with inter-orbital spin-triplet pairs
\begin{align}
p_{\rm t}^{\gm\gm'} (\bm 0) &= \sum_i c^\dg_{i\gm\ua} c^\dg_{i\gm'\ua} + {\rm H.c.}
\label{eq:3orb_pair_triplet}
\end{align}
for $J>0$ and $\gm\neq \gm'$, or intra-orbital spin-singlet pairs
\begin{align}
p_{\rm s}^{\gm} (\bm 0) &=  \sum_i c^\dg_{i\gm\ua} c^\dg_{i\gm\da} + {\rm H.c.}
\label{eq:3orb_pair_singlet}
\end{align}
for $J<0$.
Here we choose these expressions for the pair amplitudes, although we can classify them using a similar method as given in Eqs.~\eqref{eq:pair1} and \eqref{eq:pair2} using Pauli and Gell-Mann matrices.
We have the symmetry relations
$\mathscr{P}_0 p_{\rm t}^{\gm\gm'} \mathscr{P}_0^{-1} = - p_{\rm t}^{\gm\gm'}$ and
$\mathscr{P}_0 p_{\rm s}^{\gm}     \mathscr{P}_0^{-1} = - p_{\rm s}^{\gm}$.
While the signs are reversed after the transformation, i.e., the phases of the pair amplitudes are rotated by $\pi$, their forms are unchanged after the particle-hole transformation.

To understand the implications of the above symmetry relations, let us first consider the $J>0$ model at half filling.
In this case the same-spin electrons tend to occupy the same site to form Cooper pairs.
At $\Delta = 0$, the three pairs $p_{\rm t}^{12}$, $p_{\rm t}^{23}$ and $p_{\rm t}^{31}$ are degenerate.
For $\Delta >0$, as illustrated in the top part of Fig.~\ref{fig:config}, one expects that $p_{\rm t}^{12}$ is more stable than $p_{\rm t}^{23}$ and $p_{\rm t}^{31}$ because the orbitals with $\gm=1,2$ have a lower energy than $\gm=3$.
In a similar manner one may also speculate that for $\Delta < 0$ the pairs $p_{\rm t}^{23}$ and $p_{\rm t}^{31}$ are more stable than $p_{\rm t}^{12}$.
However this is {\it incorrect}. The pair $p_{\rm t}^{12}$ is the most stable even for $\Delta <0$.
This is because the pairing amplitudes are the same for $\Delta >0$ and $\Delta <0$, according to the above particle-hole symmetry argument, even though the Cooper pairs are formed in the higher-energy orbital (broken circle in the upper panel of Fig.~\ref{fig:config}) when $\Delta<0$.
This at first sight counter-intuitive result can be rationalized by considering the degeneracy: 
the $p_{\rm t}^{23}$ and $p_{\rm t}^{31}$ pairs are destabilized by fluctuations among these energetically degenerate states,
which may diminish
the pairing compared to the non-degenerate case.
On the other hand, the $p_{\rm t}^{12}$ pairs are not subject to such fluctuations and are therefore more stable.

One can also think of other effects which help explain this peculiar behavior.
Whereas the lower-energy orbital is easily occupied and  
pairs between low-energy and high-energy electrons can be formed, 
the mobility of these pairs, which is also important for realizing superconductivity, is reduced because of the high occupancy of the low-energy orbitals.
On the other hand, for higher-energy orbitals the mobility of the pairs is high, although the pairs themselves are harder to form. 
As a result, the superconducting state resulting from pairs in the high energy orbitals is actually more robust, which is also supported by the numerical results.

The same argument can also be applied to the case of $J<0$.
At $\Delta = 0$, the three pairs $p_{\rm s}^{1}$, $p_{\rm s}^{2}$ and $p_{\rm s}^{3}$ are degenerate.
For $\Delta <0$, as illustrated in the lower panel of Fig.~\ref{fig:config}, one expects that $p_{\rm s}^{3}$ is more stable than $p_{\rm s}^{1}$ and $p_{\rm s}^{2}$.
On the other hand, for $\Delta>0$ one might speculate that $p_{\rm s}^{1}$ and $p_{\rm s}^{2}$ are more stable than $p_{\rm s}^{3}$, but this is again incorrect.
Although the level of the orbital $\gm =3$ is higher than the others, the most stable Cooper pair is formed in this high-energy orbital.

Note that in the above discussion, we do not invoke any subtleties of the system, only the particle-hole symmetry of the original Hamiltonian without CF splitting.
Keeping these facts in mind, we can readily understand the result in the following subsections.

\subsection{$J>0$ case}

\begin{figure}[t]
\begin{center}
\includegraphics[width=85mm]{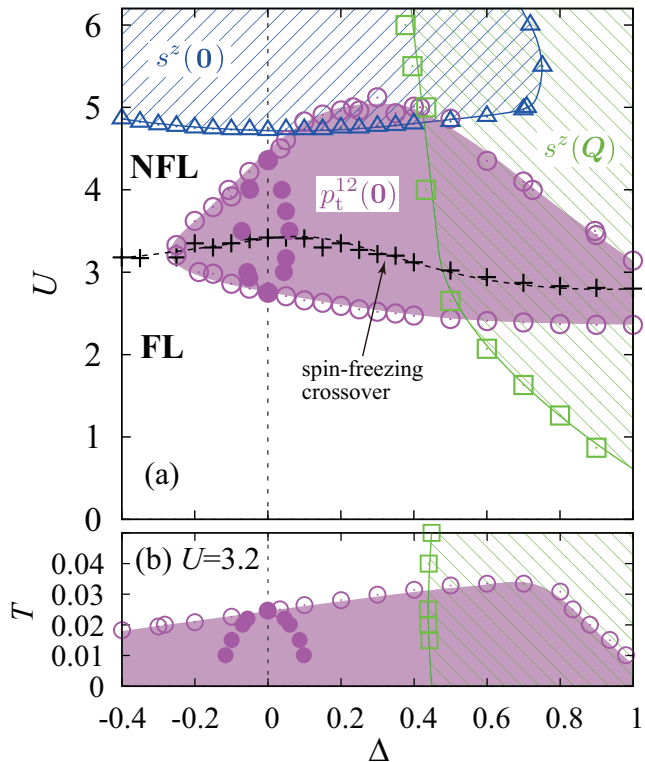}
\caption{(Color online)
(a) Phase boundaries determined by the divergence of susceptibilities for the parameters $n=2$, $J/U=1/4$ and $T=0.02$.
The limit $\Delta \rightarrow +\infty$ corresponds to the two-orbital model at half filling.
(b) Phase diagram in the temperature ($T$)-CF splitting ($\Delta$) space for $U=3.2$.
}
\label{fig:phase_PJ}
\end{center}
\end{figure}

In the remainder of this section, we show results for the three-orbital model with CF splitting, focusing again on the case with Ising anisotropy ($\al=0$). 
For $J>0$, 
the pairing amplitude is given by Eq.~\eqref{eq:3orb_pair_triplet}, and the magnetic moments for the three-orbital model are defined by
\begin{align}
s^z (\bm q) &= \sum_{i\gm\sg} c^\dg_{i\gm\sg} \sigma^z_{\sg\sg}  c_{i\gm\sg} \epn^{-\imu \bm q\cdot \bm R_i}. 
\end{align}
Other ordered states are not considered here since these are covered by the orders $p^{\gm\gm'}_{\rm t}(\bm 0)$ and $s^z(\bm q)$ in the parameter range considered in this paper.

At $\Delta=0$ the phase diagram for the three-orbital model \cite{hoshino2015} has properties similar to those of the two-orbital model shown in Fig.~\ref{fig:away_half}.
Here we discuss the effect of the CF splitting on the three-orbital model with $n=2$, because it is relevant to Sr$_2$RuO$_4$ as discussed later.
Figure~\ref{fig:phase_PJ}(a) shows the $U$-$\Delta$ phase diagram at $n=2$ and $T=0.02$.
Note that we do not have a symmetry between $\Delta>0$ and $\Delta<0$ for the case away from half-filling.
The antiferromagnetic phase ($s^z(\bm Q)$) dominates for sufficiently large $\Delta$ because in the limit $\Delta\rightarrow \infty$ the system becomes the half-filled two-orbital model discussed in the previous section.
In the large-$U$ region the ferromagnetic order ($s^z(\bm 0)$) appears.

The inter-orbital spin-triplet superconductivity is realized at intermediate values of $U$ and for a large range of $\Delta$. The spin-freezing crossover defined by the maximum of Eq.~\eqref{eq:def_local_fluct} is also plotted, and it is obvious that the superconductivity is stabilized near this line.
For $\Delta\ne 0$, the degeneracy of the three pairs is lifted, and the most stable one is $p_{\rm t}^{12}$ regardless of the sign of $\Delta$, as explained in the previous subsection. 
The other pairings $p_{\rm t}^{23,31}$ (filled circles in Fig.~\ref{fig:phase_PJ}) are quickly suppressed.

The CF splitting dependence of the transition temperature is shown in Fig.~\ref{fig:phase_PJ}(b).
It is notable that the transition temperature is enhanced by the CF splitting for $\Delta>0$.
This behavior might be due to a reduction of fluctuations among orbitals 
and an enhanced probability of pair formations
by the splitting of the degenerate orbitals.
For $\Delta < 0$ such an enhancement is not seen, presumably because the most stable Cooper pair is in this case formed in the high-energy orbitals; this pair formation is hence destabilized by the CF splitting. 
Still, the $p_\text{t}^{12}$ inter-orbital spin-triplet superconductivity can be observed in a wide region of $|\Delta|$ at low temperatures.

Our simulations 
are relevant for strontium ruthenate compounds.  
In Sr$_2$RuO$_4$, the filling is $n=4$, which is identical to $n=2$ due to particle-hole symmetry, 
and the interaction is estimated as $U\simeq 3.2$ \cite{medici2011}. 
In this compound the CF splitting among the three t$_{\rm 2g}$ orbitals has the form of Eq.~\eqref{eq:CF_def} due to the tetragonal symmetry of the crystal.
According to Figs.~\ref{fig:phase_PJ}, the spin-triplet superconductivity occurs at low temperatures even for finite CF splittings 
in the physically reasonable range. The pairing is most robust for same-spin electrons in the degenerate orbitals. 
We thus obtain consistent results for this strontium ruthenate compound, although for a more detailed comparison with experiments we should consider 
the realistic band structure.

Here we briefly comment on the effect of $\al$ which restores the isotropy.
For the present spin-triplet superconductivity with $J>0$, the effect of spin-flip terms is relevant and suppresses this pairing due to the fluctuations among the three spin-triplet states \cite{hoshino2015}.
Indeed we could not find the spin-triplet superconductivity down to the lowest accessible temperatures for $\al=1$. 
Thus the anisotropy in spin space is important for its realization.
Despite this fact, the spin-triplet superconducting state can be realized in Sr$_2$RuO$_4$,  since only a small anisotropy is necessary for its stabilization \cite{hoshino2015}.

\subsection{$J<0$ case}

\begin{figure}[t]
\begin{center}
\includegraphics[width=85mm]{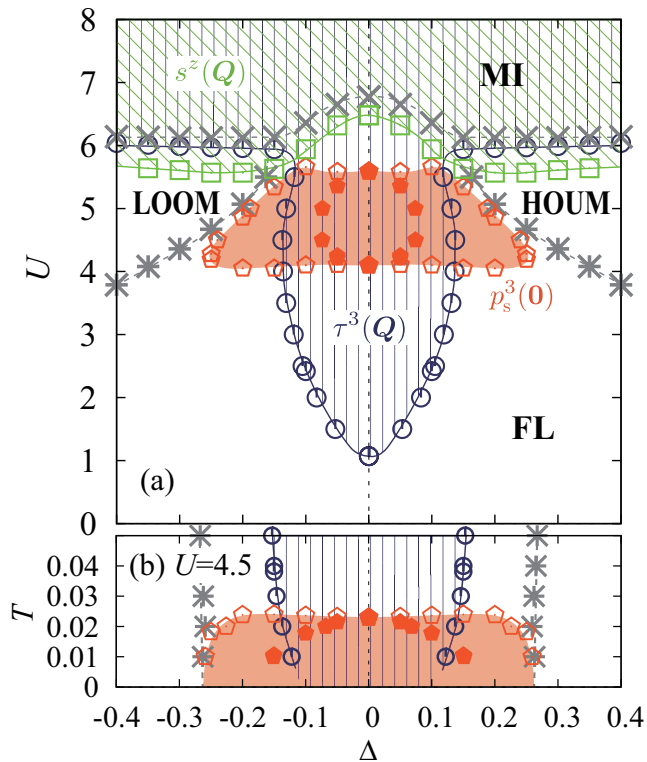}
\caption{(Color online)
(a) Phase boundaries determined by the divergence of susceptibilities for $n=3$, $J/U=-1/10$ and $T=0.02$. 
The $\Delta<0$ part is equivalent to the $\Delta>0$ part due to the particle-hole symmetry at half-filling.
Here, in addition to Fermi liquid (FL) and Mott insulator (MI) phases, we have a higher-orbital unoccupied metal (HOUM) for $\Delta >0$ and a lower-orbital occupied metal (LOOM) for $\Delta<0$.
Open (filled) pentagons correspond to $p_s^3$ ($p_s^{1,2}$) pairing. 
(b) Phase diagram in the CF splitting ($\Delta$)-temperature ($T$) space for $U=4.5$.
}
\label{fig:phase_NJ}
\end{center}
\end{figure}

Because of the relevance for fulleride compounds, it is also interesting to consider the half-filled ($n=3$) three orbital model with negative Hund coupling \cite{nomura2015}.
In addition to the magnetic moment defined in the last subsection, we consider two kinds of orbital moments defined by
\begin{align}
\tau^{3,8} (\bm q) &= \sum_{i\gm\sg} c^\dg_{i\gm\sg} \lambda^{3,8}_{\gm\gm}  c_{i\gm\sg} \epn^{-\imu \bm q\cdot \bm R_i},
\end{align}
where $\lambda^3={\rm diag}(1,-1,0)$ and $\lambda^8={\rm diag}(1,1,-2)/\sqrt{3}$ are Gell-Mann matrices.
These orbital orders are degenerate for $\Delta = 0$, but become different in the presence of CF splitting.
For the parameter range considered in this paper, the orbital order with $\tau^3(\bm Q)$ is more stable than the other, so we plot only the phase boundaries for $\tau^3(\bm Q)$.

Figure \ref{fig:phase_NJ}(a) shows the $U$-$\Delta$ phase diagram at $J/U=-1/10$ and $T=0.02$.
Although this ratio of $J/U$ is much larger than the estimate for alkali-doped fullerides \cite{nomura2012}, it is suitable for clarifying the qualitative behaviors originating from the negative Hund coupling.
Without any long-range order, we have three kinds of states depending on the parameters.
In the small and large $U$ regions, metallic and Mott insulating states are realized, respectively. 
For sufficiently large CF splitting, a new state appears in-between, which is a metallic state with a completely empty (occupied) orbital for $\Delta>0$ ($\Delta<0$) \cite{footnote_jpos}.

Now we discuss the ordered states. 
The orbital degrees of freedom are activated by the negative Hund coupling and are ordered in a staggered manner as shown in Fig.~\ref{fig:phase_NJ}(a). 
Orbital order with $\tau^3(\bm Q)$ moments is the dominant phase near $\Delta=0$.
For large $U$, antiferromagnetism is also stabilized, which is explained by the still remaining active spin degrees of freedom 
in the half-filled system.

As shown in Fig.~\ref{fig:phase_NJ}, the intra-orbital spin-singlet pairing with $p_{\rm s}^3$ is realized in the intermediate $U$ region.
This pairing is also caused by the negative Hund coupling, which favors doubly occupied orbitals.  
The other pairings with $p_{\rm s}^{1,2}$ are less stable (see filled pentagons in the figure) as discussed before.
For small $\Delta$, this superconducting state is covered by the orbital order.
With increasing CF splitting, we find a region between the orbital ordered phase and the occupied/unoccupied orbital metal, where the $p_{\rm s}^3$ pairing state is most stable.
The transition temperature as a function of $\Delta$ is plotted in Fig.~\ref{fig:phase_NJ}(b).
It is notable that the transition temperature is slightly enhanced by the CF splitting.
This enhancement can be intuitively interpreted again as 
resulting from the suppression of fluctuations among orbitals and increasing probability of pair formation.
At the same time, for $\Delta<0$ the mobility of the pairs is reduced, so that the $T_c$ is almost independent of $\Delta$. 
For $\Delta > 0$, on the other hand, the probability of pair formation instead of the mobility of pairs is decreased, to give the exactly same result as in the $\Delta < 0$ case.
The $p_\text{s}^{3}$ pairing state disappears when orbital $\gamma=3$ becomes fully occupied or empty.

Finally we discuss the effect of the parameter $\al$. 
In the $J<0$ case, the pair hopping is relevant and the spin-flip is not. 
According to Ref.~\onlinecite{nomura2015}, the pair hopping substantially enhances the intra-orbital spin-singlet pairing. Hence the transition temperature should increase for $\al>0$, 
which is in contrast to the $J>0$ case. 
While the realistic value of $J$ is tiny ($J/U\sim -0.025$ for fullerides \cite{nomura2012}), the pairing from negative Hund coupling has this advantage and has a chance to be realized in systems with anisotropic interaction.
The effect of the increasing orbital fluctuations on the intra-orbital spin-singlet superconducting state for $\al>0$, however, remains to be investigated.

\section{Summary and Discussion}

We have presented a systematic analysis of the electronic ordering instabilities in two- and tree-orbital Hubbard models. We have used DMFT in combination with a numerically exact hybridization expansion impurity solver and a semi-circular density of states. Our results thus represent the generic phase diagrams of high-dimensional lattice models, irrespective of details of the bandstructure. Only uniform and staggered order parameters have been considered, and we focused on the intermediate coupling regime, which is relevant for most unconventional multi-band superconductors and cannot be accessed reliably by approximate methods such as the fluctuation-exchange approximation.

In the case with positive Hund coupling, we found instabilities to antiferromagnetic and ferromagnetic orders, spin-orbital order and orbital-singlet spin-triplet superconductivity. 
In the model with negative Hund coupling we identified antiferro orbital order, antiferromagnetism, and intra-orbital spin-singlet pairing. We have shown how symmetry relations can be used to connect some of these ordered phases, and to understand the Cooper pairings in three-orbital models with CF splittings. 

In particular, we showed that in the two-orbital model with $J/U=2/5$, the orbitally degenerate system away from half-filling can be mapped exactly onto the half-filled system with CF splitting, and that the orbital-singlet spin-triplet superconductivity in the former system corresponds to spin-orbital order in the latter. The qualitative correspondence between these at first sight different physical situations can be expected to hold in a wider parameter regime. Since the fluctuating local moments at the border of the spin-frozen metal regime play a crucial role in stabilizing the spin-triplet superconducting state, and the spin moments are unaffected by the mapping, this symmetry argument also implies that fluctuating moments drive the instability to spin-orbital order.
Thus the diagonal and offdiagonal orders can be understood in a unified manner.

Crystal field splittings of the ``2/1" type in the three orbital model can have a surprising effect on the superconducting state. In the spin-triplet state with filling $n=2$ and $J>0$, the most stable pairs are always formed in the two degenerate orbitals, even if the energy of these orbitals is increased by the CF. On the other hand, in the half-filled spin-singlet state with $J>0$, the most stable pairs are always formed in the non-degenerate orbital, irrespective of the sign of the CF. We explained this behavior using a particle-hole transformation, and also gave an intuitive interpretation based on fluctuations of the condensate and mobility of Cooper pairs.

The use of symmetry relations is also interesting from a computational point of view, since the rotational invariance of the Slater-Kanamori interaction allows to express certain four-point correlation functions in terms of others which can be easily measured in standard hybridization expansion Monte Carlo simulations \cite{gull2011}. This trick however fails as soon as the rotational invariance is broken, e.g. by the presence of CF splittings. Whether or not it is possible to extend this technique to more general situations by exploiting additional symmetries is an interesting open problem. 
An alternative strategy is to directly measure all types of four-point correlation functions using a worm sampling algorithm \cite{gunacker2015}.

Our three-orbital calculations with filling $n=2$ (equivalent to $n=4$) can be regarded as toy model simulations of Sr$_2$RuO$_4$. This material is a quasi-2D system with tetragonal symmetry and has a ``2/1'' type CF splitting.
While the CF splitting quickly suppresses the superconductivity in the non-degenerate orbital,
the pairing between the degenerate orbitals is robust. 
Our calculations are consistent with a fluctuating-moment induced spin-triplet superconductivity in this compound, which has a partial orbital degeneracy ($zx$ and $yz$ orbitals).  
More accurate simulations would need to take into account the realistic bandstructure.

The half-filled three-orbital model with negative Hund coupling is of interest in connection with fulleride compounds, where the Jahn-Teller screening of the small $J$ leads to a stabilization of the low-spin states \cite{capone2009}.
The effect of CF splitting is to further stabilize the spin-singlet pairing in the non-degenerate orbital,
and to suppress a competing orbital ordered phase. 
While the $J$ parameters used in our study are larger than the ab-initio estimates \cite{nomura2012}, this finding has some implications on the recently reported light-induced superconductivity in K$_3$C$_{60}$ \cite{mitrano2015}. One possible explanation for the enhanced $T_c$, proposed in Ref.~\cite{mitrano2015}, is that the driving of a phonon of $T_{1u}$ symmetry leads to an essentially static distortion of the C$_{60}$ molecules, and hence a splitting of the initially degenerate molecular orbitals. Our result shows that this splitting, if studied within an equilibrium formalism, 
can contribute to a slightly enhanced pairing. The related effect of orbital differentiation in the interaction parameters still needs to be systematically explored.
Given the large enhancement of $T_c$ observed in the experiments,
one may speculate that the $J$ parameter is effectively modified in the driven state (enhanced dynamical Jahn-Teller effect).
This, and other non-equilibrium phenomena should be investigated within a Floquet formalism
or using the non-equilibrium extension of DMFT \cite{aoki2014}.

\acknowledgements

S.H. acknowledges financial support from JSPS KAKENHI Grant No. 13J07701 and P.W. support from FP7 ERC starting grant No. 278023. The authors benefited from the Japan-Swiss Young Researcher Exchange Program 2014 coordinated by JSPS and SERI.
The numerical calculations have been performed on the BEO04 cluster at the University of Fribourg and the supercomputer at ISSP (University of Tokyo).

\appendix

\section{Bethe Lattice in Infinite Dimensions}
\label{appendixBethe}

Here we consider some properties of the Bethe lattice in infinite dimension.
Since the wave vectors are ill-defined in this special lattice, we have to deal with real space.
However, we will show that the concept of wave vectors can be partially applied, and that a treatment similar to that of ordinary lattices is possible.

\subsection{Single-Particle Green Function}
Let us here consider the tight-binding model for non-interacting spinless fermions on the Bethe lattice:
\begin{align}
\mathscr{H}_0 = -t \sum_{\la ij \ra} c^\dg_i c_j,
\end{align}
where the summation with respect to the site index is over the nearest neighbor pairs.
We begin with the equations of motion given by
\begin{align}
-\partial_\tau c_i &= -t \sum_{\delta} c_{i+\delta},
\\
-\partial_\tau c_{i+\delta} &= -t c_i -t \sum_{\delta+\delta'\neq 0} c_{i+\delta+\delta'},
\end{align}
where $-\partial_\tau O=[O,\mathscr{H}_0]$.
The site index $i+\delta$ denotes the nearest neighbor sites of the site $i$.
The Fourier transformation of the equations of motion for the Green function defined by $G_{ij}(\tau) = - \la T_{\tau} c_i (\tau) c_j^\dg \ra$ is given by
\begin{align}
z G_0(z) &= 1 - td G_1(z),
\\
z G_k(z) &= -t G_{k-1}(z) - td G_{k+1}(z),
\end{align}
for $k=1,2,\cdots$, which represents the number of sites between the positions at
$i$ and $j$.
Here $z$ is a complex frequency, $d$ is a connectivity, and $G_0$ is a local Green function.
We have taken the limit $d\rightarrow \infty$.
Defining $\al_k = G_k/G_{k-1}$, one can show the following relation 
\begin{align}
\al_k &= \frac{-t}{z+td\al_{k+1}} 
= \frac{-t}{z+td \frac{-t}{z+td \frac{-t}{z+\cdots}}}
= \frac{-t}{z+td\al_{k}}.
\end{align}
Hence the ratio $\al_k$ is independent of $k$, and we can write it as $\al_k = \al$.
This quantity is explicitly derived by solving the quadratic equation and we obtain the Green function as
\begin{align}
G_k(z) = - \al(z)^{k+1} / t
. \label{eq:green_bethe}
\end{align}
One of the two solutions for $\alpha$ is chosen so that it behaves as $G_0(z) \rightarrow 1/z$ when $|z|\rightarrow \infty$.
The form of the Green function is the same as the one obtained in Ref.~\onlinecite{eckstein2005}.
For interacting systems, the local self-energy $\Sigma$ is included by the replacement $z\longrightarrow z-\Sigma$.

\subsection{Two-Particle Green Function}
We define the two-particle Green functions without vertex parts by
\begin{align}
\chi^0_{\rm unif}(z) &= - \frac 1 N \sum_{ij}  G_{ij}(z) G_{ji}(z),
\\
\chi^0_{\rm stag}(z) &= - \frac 1 N \sum_{ij} \lambda_i \lambda_j G_{ij}(z) G_{ji}(z),
\end{align}
which are relevant to uniform and staggered susceptibilities, respectively.
Here $\lambda_i=\pm 1$ is a sign which depends on the sublattice.
Substituting the results in the previous subsection, we obtain
\begin{align}
\chi^0_{\rm unif} = - \frac{(\al/t)^2}{1-d\al^2}
, \hspace{3mm}
\chi^0_{\rm stag} = - \frac{(\al/t)^2}{1+d\al^2}.
\end{align}
This leads to the simpler expressions 
\begin{align}
\chi^0_{\rm unif} 
= \frac{\diff G_0}{\diff z}
,\hspace{3mm}
\chi^0_{\rm stag} 
= - \frac{G_0}{z}.
\label{eq:unif_stag}
\end{align}
Thus we derive the uniform and staggered components of two-particle Green functions.

We now consider another representation of the above two-particle Green functions, and introduce a density of states and a ``wave-vector $\bm k$ summation'' by
\begin{align}
G_0(z) = \int \frac{\rho(\ep)}{z-\ep} \diff \ep
\equiv \frac 1 N \sum_{\bm k} g_{\bm k}(z)
, \label{eq:def_k_sum}
\end{align}
where $\rho(-\ep) = \rho(\ep)$, $g_{\bm k}(z) = 1/(z-\ep_{\bm k})$ and $N=\sum_{\bm k}1$.
Here we do not have to know the specific form of $\ep_{\bm k}$, and only assume the existence of a vector $\bm Q$ defined by $\ep_{\bm k+\bm Q} = -\ep_{\bm k}$.
We also introduce the $\bm q$-dependent two-particle Green function by
\begin{align}
\tilde \chi^0_{\bm q} (z) &= -\frac 1 N \sum_{\bm k} g_{\bm k}(z) g_{\bm k+\bm q}(z)
.
\end{align}
Using Eq.~\eqref{eq:def_k_sum}, one can show the relations
$\tilde \chi^0_{\bm 0} = \chi^0_{\rm unif}$ and
$\tilde \chi^0_{\bm Q} = \chi^0_{\rm stag}$.
Thus the pseudo wave vectors $\bm q=\bm 0$ and $\bm q=\bm Q$ correspond to the uniform and staggered components, respectively.
Namely, the Bethe lattice can be partly handled as if it were a simple cubic lattice which has the same relation $\ep_{\bm k+\bm Q} = -\ep_{\bm k}$ with the staggered ordering vector $\bm Q = (\pi,\pi,\pi)$.
Note that this analogy is valid only for the uniform and staggered components.

\section{Physical Interpretation of Order Parameters}
\label{appendix_eg}

Let us physically interpret the orbital and spin-orbital moments in terms of doubly degenerate e$_{\rm g}$ orbitals for $d$ electrons under the cubic symmetry.
The wave functions are written as
\begin{align}
| \gm=1 \ra &= |0\ra ,
\\
| \gm=2 \ra &= (|2\ra + |-2\ra ) /\sqrt 2 ,
\end{align}
where $|m\ra$ in the right-hand side is the eigenstate of the angular momentum $\ell_z$.
With use of this expression, we can show the relations
\begin{align}
\tau^z &\propto 3 \ell_z^2 - \bm \ell^2,
\\
\tau^x &\propto \ell_x^2 - \ell_y^2,
\\
\tau^y &\propto \overline{\ell_x \ell_y \ell_z},
\end{align}
where the overline symmetrizes the product of operators as e.g. 
$\overline{\ell_x \ell_y} = (\ell_x \ell_y + \ell_y \ell_x)/2!$.
This can be shown by standard quantum mechanics calculations for angular momentum.
Thus $\tau^z$ and $\tau^x$ are rank 2 operators, while $\tau^y$ is a rank 3 operator.
For the spin-orbital moment, we further put the spin moment on the orbital moment $\tau^\mu$, and hence the rank of the operator is increased by one.
Namely, $o^{z\mu}$ and $o^{x\mu}$ are rank 3 operators, and $o^{y\mu}$ is a rank 4 operator.
In the context of $f$-electron systems, rank 0,1,2,3,4 operators are called `scalar (monopole)', `dipole', `quadrupole', `octupole' and `hexadecapole', respectively \cite{kuramoto2009}.
The odd (even) rank tensor is time-reversal odd (even).


\begin{thebibliography}{99}
\bibitem{footnote_jpos} A similar orbital occupied/unoccupied state also appears in the $J>0$ case. It is however out of the parameter range shown in Fig.~\ref{fig:phase_PJ}.
\end{thebibliography}
\end{document}